\documentclass[aps,pra,twocolumn,reprint,amsmath,amssymb,superscriptaddress,floatfix,footinbib,longbibliography]{revtex4-2}

\usepackage{epsfig}
\usepackage{graphicx}
\usepackage{epstopdf}

\usepackage[T1]{fontenc}
\usepackage[applemac]{inputenc}
\usepackage{units}
\usepackage{comment}

\usepackage{amsmath,amssymb,natbib,bm}
\usepackage{psfrag}

\usepackage{amsthm}
\usepackage{bbm}

\usepackage{tikz}
\usetikzlibrary{arrows}

\usepackage{color}
\usepackage{url}

\usepackage[colorlinks]{hyperref}
\hypersetup{%
        plainpages=true,
        breaklinks=true,
        hypertexnames=false,
        pageanchor=true,
        colorlinks=true,
        linkcolor={blue},
        citecolor={magenta},
        urlcolor={blue},
        anchorcolor={black}
      }
\usepackage[all]{hypcap} % let hyperlinks correctly point to figures rather than their captions; must be loaded after the hyperref package!

\usepackage{mleftright} %to get rid of the extra space around parentheses with \left and \right: write \mleft, \mright instead 

\newcommand{\figref}[1]{\mbox{Fig.~\ref{#1}}}
\newcommand{\tabref}[1]{\mbox{Table~\ref{#1}}}
\newcommand{\secref}[1]{\mbox{Sec.~\ref{#1}}}

\newcommand{\appref}[1]{\mbox{Appendix~\ref{#1}}}
\renewcommand{\eqref}[1]{\mbox{Eq.~(\ref{#1})}}

\newcommand{\figpanel}[2]{Fig.~\hyperref[#1]{\ref*{#1}(#2)}}
\newcommand{\figpanels}[3]{Fig.~\hyperref[#1]{\ref*{#1}(#2)-(#3)}}
\newcommand{\figpanelNoPrefix}[2]{\hyperref[#1]{\ref*{#1}(#2)}}
\newcommand{\ket}[1]{\mleft|#1 \mright \rangle}

\newcommand{\ketbra}[2]{\mleft| #1 \rangle \langle #2 \mright|}

\newcommand{\be}{\begin{equation}}
\newcommand{\ee}{\end{equation}}
\newcommand{\bea}{\begin{eqnarray}}
\newcommand{\eea}{\end{eqnarray}}

\begin{document}

\title{Detecting quantum speedup of random walks with machine learning}

\date{\today}

\author{Hanna Linn}
\thanks{Equal author contributions}
\email{hannlinn@chalmers.se}
\affiliation{Department of Microtechnology and Nanoscience, Chalmers University of Technology, 412 96 Gothenburg, Sweden}

\author{Yu Zheng}
\thanks{Equal author contributions}
\email{zhyu@chalmers.se}
\affiliation{Department of Microtechnology and Nanoscience, Chalmers University of Technology, 412 96 Gothenburg, Sweden}

\author{Anton Frisk Kockum}
\email{anton.frisk.kockum@chalmers.se}
\affiliation{Department of Microtechnology and Nanoscience, Chalmers University of Technology, 412 96 Gothenburg, Sweden}

\begin{abstract}

We explore the use of machine-learning techniques to detect quantum speedup in random walks on graphs. Specifically, we investigate the performance of three different neural-network architectures (variations on fully connected and convolutional neural networks) for identifying linear, cyclic, and random graphs that yield quantum speedups in terms of the hitting time for reaching a target node after starting in another node of the graph. Our results indicate that carefully building the data set for training can improve the performance of the neural networks, but all architectures we test struggle to classify large random graphs and generalize from training on one graph size to testing on another. If classification accuracy can be improved further, valuable insights about quantum advantage may be gleaned from these neural networks, not only for random walks, but more generally for quantum computing and quantum transport.

\end{abstract}

\maketitle

%%%%%%%%%%%%%%%%%%%%%%%%%%%%%%%%%%%%%%%%%%%%%%%

\section{Introduction}

Computations using quantum systems~\cite{Nielsen2000} have the potential to offer significant speedup over classical computers for certain problems~\cite{Georgescu2014, Montanaro2016, Wendin2017, Preskill2018, McArdle2020, Bauer2020, Cerezo2021, Cerezo2022}. Indeed, some recent experiments with quantum-computing devices have challenged the limits and performances of classical computing hardware~\cite{Arute2019, Zhong2020, Wu2021a, Zhu2022, Madsen2022, Kim2023}. However, it can be difficult to find out whether a quantum computer can provide a meaningful speedup for solving a certain problem~\cite{Aaronson2022, Ronnow2014, Tang2019, Babbush2021, Tang2022}. In many cases, it also remains an open question what quantum resources and properties of a quantum-computing device are necessary to enable a quantum advantage~\cite{Chitambar2019, Nielsen2000, Bravyi2005, Knill2005, Mari2012, Albarelli2018, Garcia-Alvarez2021, Calcluth2023, Zheng2023}.

In this article, we focus on quantum versions of random walks~\cite{Aharonov1993, Kempe2003, Ambainis2003, Kendon2007, Mulken2011, Venegas-Andraca2012, Xia2020, Kadian2021} and how to determine when they can provide an advantage over classical random walks. The classical random walk~\cite{Pearson1905, Spitzer1964} has many applications~\cite{Kac1947, Nosofsky1997, Brin1998, Grady2006, Gkantsidis2006, Codling2008, Xia2020}, both in algorithms and in modelling of physical systems. This motivated studies of quantum walks~\cite{Childs2003, Chakraborty2016, Apers2021, Kendon2020, Ambainis2020, Chakraborty2020}, which have turned out to be able to outperform classical random walks in some instances (usually, by traversing a graph faster). It has also been shown that quantum walks can perform universal quantum computation~\cite{Childs2009, Childs2013}. As such, improving the understanding of when quantum walks outperform classical random walks may give clues about when quantum computers can outperform classical computers. Recent experiments have demonstrated implementations of quantum walks on tens, hundreds, and even thousands of nodes using several hardware platforms~\cite{Tang2018, Chen2018, Yan2019, Gong2021, Young2022}.

Determining when quantum walks are faster than classical random walks can be done analytically or by simulating both walks on a graph. While there are analytical results for certain types of graphs~\cite{Ambainis2001, Aharonov2001, Childs2003, Kempe2005, Solenov2006, Krovi2006, Santos2010, Makmal2014, Balasubramanian2023} (and progress toward more generally applicable criteria~\cite{Gualtieri2020, Chakraborty2020}), the space of possible graphs is vast. Simulations can handle any type of graphs, but only provide answers for individual cases without giving information about general properties. As the graph size and connectivity grow, the simulation time also grows; the propagation time for the walker usually scales polynomially with the size of the graph~\cite{Lawler1986, Aldous1999}.

Here, we instead apply machine-learning methods to classify on which graphs a quantum walker will reach its target faster than a classical walker. Machine learning has in the past few years been applied to aid quantum technologies in many ways~\cite{Krenn2023}, e.g., for quantum error correction~\cite{Torlai2017, Andreasson2019, Fitzek2020}, quantum state and process tomography~\cite{Torlai2018, Flurin2020, Ahmed2021, Ahmed2021a, Gebhart2023}, quantum control~\cite{Fosel2018, Niu2019, Wittler2021}, compilation of quantum circuits~\cite{Fosel2021, Moro2021, Pozzi2022}, and more~\cite{Krenn2016, Carleo2017, Krenn2020, Bharti2020}. Some authors have used machine learning to identify non-classicality~\cite{Gebhart2020, Harney2020} or to estimate the capability of a quantum computer~\cite{Hothem2023}. In particular, for quantum walks, there have been two initial studies of how neural networks can classify whether a quantum walker has an advantage over a classical one on various graphs~\cite{Melnikov2019, Melnikov2020}.

We go beyond the work in Refs.~\cite{Melnikov2019, Melnikov2020} by exploring several different neural-network architectures, larger data sets, and more carefully crafted data sets, in order to increase the accuracy of classifying which graphs yield an advantage for a quantum walker. We also explore how the model changes when the quantum walker is initialized in a superposition state, motivated by the fact that universal quantum computing is possible with a limited set of quantum gates provided that the initial state is allowed to be more complicated than, e.g., the ground state or a coherent state~\cite{Bravyi2005, Knill2005, Albarelli2018}.  If the accuracy of a neural-network classifier for quantum walks becomes high enough, it would be an appealing alternative to simulations for large general graphs, since the trained neural network then quickly could determine whether there is a quantum advantage. A high-accuracy neural-network classifier could also yield important insights about criteria (e.g., length of the graph or symmetries such as cycles) that can be used to assess when quantum advantage occurs. 

We find that the neural-network architecture in Ref.~\cite{Melnikov2019} performs at least as well as the fully connected and convolutional networks that we compare it to and improves its performance compared to before after being fed more balanced training data. However, the task of classifying larger random graphs, according to whether a quantum walker is faster than a classical random walker or not, is highly challenging for all three neural-network architectures. To distinguish between quantum walks starting in different states turns out to be even more difficult. Likewise, for a neural network to generalize to larger graph sizes after only having been trained on smaller ones is also very hard.

Our results add to the literature on identifying quantum advantages. If the performance of these neural networks, or other neural-network architectures, can be further improved, there may be much to learn about quantum advantages from them. Such learnings are not necessarily limited to random walks; they may also apply to other ways of quantum information processing or to quantum transport.

This article is organized as follows. In \secref{sec:Theory}, we provide a theoretical background on classical random and quantum walks, showing how they are defined and simulated. We also give some background on the superposition states known as ``magic states''. In Section~\ref{sec:Methods}, we describe the methods we used in our study, including the neural-network architectures and the datasets used. The main results of our numerical experiments are presented in Section~\ref{sec:Results}. In Section~\ref{sec:ConclusionOutlook}, we discuss our findings and provide an outlook for future research. Additional details of the methods and results are provided in appendixes.

%%%%%%%%%%%%%%%%%%%%%%%%%%%%%%%%%%%%%%%%%%%%%%%

\section{Theory for classical and quantum walks}
\label{sec:Theory}

The random walks we study in this article are continuous-time random walks (CTRWs). In this section, we introduce classical CTRWs and their quantum versions, continuous-time quantum walks (CTQWs). For CTQWs, we further show how to formulate them when the initial state is a superposition state, and we also discuss how dark states can affect their dynamics.

%%%%%%%%%%%%%%%%%%%%%%%%%%%%%%%%%

\subsection{Continuous-time random walk}
\label{sec:CTRW}

%---------------------------------------------------------------------------------%
\begin{figure}
\includegraphics[width = \linewidth]{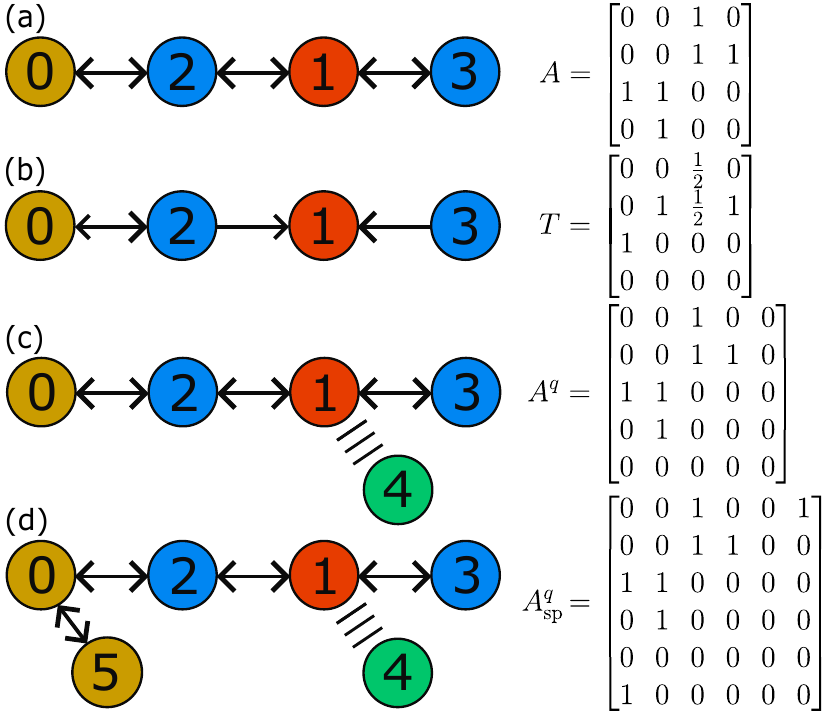}
\caption{Setups for classical and quantum walks on a linear graph with four nodes $V=\{0, 1, 2, 3\}$ and three edges $E = \{(0,2), (2,1), (1,3)\}$. The initial node is yellow, the target node is red, ordinary nodes are blue, and an auxiliary sink node is green.
(a) The undirected original graph together with its symmetric adjacency matrix $A$ on the right.
(b) The directed graph for a CTRW catches the particle in the target node. Its transition matrix $T$ is shown on the right.
(c) For the CTQW, an unconnected auxiliary node is added, where measurements are done instead of at the target node. The corresponding adjacency matrix $A^q$ is shown on the right.
(d) To initialize the CTQW in a superposition state, we add another starting node, extending the graph with an additional edge $(0,5)$, yielding the adjacency matrix $A^q_{\rm sp}$ on the right.}
\label{fig:adjacency_graph_matrix}
\end{figure}
%---------------------------------------------------------------------------------%

A node-centric CTRW on an undirected graph $G(E,V)$, where $E$ is the set of edges and $V$ is the set of vertices (nodes) of the graph, starts in an initial node, which we label $i=0$, and ends in a target node, which we label $i=1$. Which nodes a particle walking on the graph can move between is given by the adjacency matrix $A$, which is of size $n \times n$ when $|V|=n$. If the element $a_{ij}$ of $A$ equals one, the particle can move to node $i$ from node $j$ along an edge. See \figpanel{fig:adjacency_graph_matrix}{a} for an example with four nodes in a line graph.

Since we here want to catch the particle in the target node, we modify $A$ from being symmetric to
\begin{equation}
A_{ij}^c =
\begin{cases}
1, & \text{if } (i, j) \in E \land i \not = 1, \\
1, & \text{if } i = j = 1, \\
0, & \text{otherwise},
\end{cases} 
\label{eq:A_c}
\end{equation}
i.e., a particle can move to the target node, but not from it. See \figpanel{fig:adjacency_graph_matrix}{b} for how this modifies the graph compared to the example in \figpanel{fig:adjacency_graph_matrix}{a}.

The probability of finding the particle in node $i$ at time $t$ is given by the probability distribution $p_i(t)$. 
The master equation describing the time evolution of the vector $p(t)$ of node occupation probabilities for a node-centric CTRW on a graph is
\begin{equation}
\frac{dp(t)}{dt} = (T-I) p(t),
\label{eq:node_centric_diff}
\end{equation}
where $I$ is the identity matrix, and $T$ is the transition matrix with elements
\begin{equation}
T_{ij} = \frac{A_{ij}^c}{\text{degree}(j)}.
\label{eq:tranisionM}
\end{equation}
See \figpanel{fig:adjacency_graph_matrix}{b} for the $T$ connected to the graph there. The solution to \eqref{eq:node_centric_diff} is
\begin{equation}
p(t) = e^{(T-I) t} p(0),
\label{eq:prob_c}
\end{equation} 
where $p(0) = [1, 0, \ldots, 0]$ since the probability of finding the particle in the initial node at time $t=0$ is one~\cite{Masuda_2017, Aldous1999}.

The time it takes for a particle to find the target node is called the hitting time. This time can be found by simulating \eqref{eq:prob_c}.
More generally, the mean hitting time is defined as the time $\tau$ when the probability $p_1(\tau)$ of finding the particle in the target node is above the threshold $p_{\rm th}$ defined by the length of the characteristic path.
For random graphs, the characteristic path length is $p_{\rm th} = \frac{1}{\log n}$~\cite{log_n_watts_2004}.

%%%%%%%%%%%%%%%%%%%%%%%%%%%%%%%%%

\subsection{Continuous-time quantum walk}
\label{sec:CTQW}

If the particle walking on the graph is a quantum particle instead of a classical particle, the walk will have a different probability distribution and a different hitting time on the same graph, thanks to quantum effects like interference~\cite{Aharonov1993}. A CTQW on the graph $G(E,V)$ can be defined with a Hilbert space $\mathcal{H}_{G(E,V)} = \{\ket{0}, ..., \ket{n-1}\}$~\cite{Farhi1998, Mulken2011} and a Hamiltonian $H = \hbar A$.

To avoid destroying the quantum state or missing the particle when measuring the target node, a sink node is connected to the target node.
Serving essentially the same purpose as cutting off the connections out from the target node for the CTRW in \secref{sec:CTRW}, the sink catches the particle by decay from the target node. The probability of hitting the target can then be measured in the sink node instead~\cite{Melnikov2020a}. The adjacency matrix is altered to include the absorbing mechanism, but the sink node does not have any edges; $A$ is padded with zeros to become an $(n+1) \times (n+1)$ matrix with elements
\begin{equation}
A_{ij}^q =
\begin{cases}
1, & \text{if } (i, j) \in E, \\
0, & \text{if } i = n \lor j = n, \\
0, & \text{otherwise}.
\end{cases} 
\label{eq:A_q}
\end{equation}
See \figpanel{fig:adjacency_graph_matrix}{c} for an example.

The probability distribution of finding the quantum particle in the node $i$ at time $t$ is given by $\rho_{ii}(t) = p_i(t)$, where $\rho(t)$ is the density matrix describing the total quantum state in the Hilbert space. The dynamics of the CTQW can be described by the Lindblad equation~\cite{Lindblad1976}
\begin{equation}
\frac{\mathrm{d} \rho(t)}{\mathrm{d} t} = - \frac{i}{\hbar} \mleft[ H , \rho(t) \mright] + \gamma \mleft( L \rho(t) L^\dag - \frac{1}{2} \mleft\{ L^\dag L, \rho(t) \mright\} \mright),
\label{eq:lindbladian}
\end{equation}
where $L = \ketbra{n}{1}$ is the jump operator yielding the decay from the target to the sink node with the rate $\gamma$~\cite{Melnikov2019}. To find $p(t)$, we solve \eqref{eq:lindbladian} with the initial state $\rho(0) = \ketbra{0}{0}$. The mean hitting time $\tau$ is then obtained by observing the population in the sink node and comparing it to the threshold probability $p_{\rm th}$.

%%%%%%%%%%%%%%%%%%%%%%%%%%%%%%%%%

\subsection{Superposition states and magic}
\label{sec:Magic}

We also consider CTQWs initialized in superposition states. To be able to do so, we add another starting node to the setup described in \secref{sec:CTQW}. The node is added similarly to the sink node, but with a bidirectional connection to the initial node. This means that we pad $A$ further with zeros, making it an $(n+2) \times (n+2)$ matrix, but add ones in the first row of the last column and in the last row of the first column:
\begin{equation}
A_{ij, \text{sp}}^q =
\begin{cases}
1, & \text{if } (i, j) \in E, \\  
0, & \text{if } i = n \lor j = n, \\
1, & \text{if } (i = 0, j = n+1) \lor (i = n+1, j = 0), \\
0, & \text{otherwise}.
\end{cases} 
\label{eq:Aq_superpos}
\end{equation}
See \figpanel{fig:adjacency_graph_matrix}{d} for an example. It is possible to imagine other ways to initialize the CTQW in a superposition state, but that is outside the scope of this article.

Including CTQWs initialized in superposition states is motivated by the fact that certain superposition states are known to be powerful in digital quantum computing. The Gottesman-Knill theorem~\cite{Gottesman1999} states that quantum circuits initialized in stabilizer states (the eigenstates of the Pauli matrices $\sigma_x$, $\sigma_y$, and $\sigma_z$; i.e., the states at $\pm 1$ of the $x$, $y$, and $z$ axes of the Bloch sphere), containing only operations from the Clifford group, and having measurements performed only in the Pauli basis, can be efficiently simulated by a classical computer. However, if one can add preparation of so-called magic states, i.e., states outside the octahedron spanned by the stabilizer states on the Bloch sphere, to such a quantum circuit, it can implement universal quantum computation~\cite{Bravyi2005, Knill2005}.

The further away from the stabilizer states the magic state is, the more resourceful it is for quantum computation. The states with the most ``magic'' are the T states, defined by
\begin{equation}
\ketbra{T}{T} = \frac{1}{2} \left[ I + \frac{1}{\sqrt{3}} (\sigma_x + \sigma_y + \sigma_z) \right],
\end{equation}
where $I$ is the identity matrix. Eight such states are located on the Bloch sphere, above the middle of the faces of the octahedron~\cite{Bravyi2005}. The specific state
\begin{equation}
\ket{T} = \cos \beta \ket{0} + e^{i\pi/4} \sin \beta \ket{1}, \quad \cos 2\beta = \frac{1}{\sqrt{3}}
\label{eq:T_state_ket}
\end{equation}
is an eigenstate of the unitary operator
\begin{equation}
T = \frac{e^{i\pi/4}}{\sqrt{2}} \begin{bmatrix} 1 & 1 \\ i & -i\\ \end{bmatrix}
\end{equation}
that implements a T gate. We use this state as an initial state for some CTQWs in this article.

%%%%%%%%%%%%%%%%%%%%%%%%%%%%%%%%%

\subsection{Influence of dark states}
\label{sec:dark}

In atomic physics, there are examples of dark states for atoms or molecules, i.e., quantum states that cannot absorb an incident photon (or emit a photon)~\cite{Dicke1954, Lenz1993}. In a similar fashion, there can be dark states in quantum walks due to quantum interference between different paths in the graph, making the quantum walker never reach the target node~\cite{Thiel2020}. This is the flip side of quantum interference sometimes speeding up quantum walks compared to classical walks.

From the understanding of dark states, it can be shown that the upper bound of $p_1 (t \rightarrow \infty) = p_{\rm det}$, the total probability that the particle eventually is detected in the target state, is controlled by $n_{\rm init}$, the number of states equivalent (with respect to the detection) to the initial state~\cite{Thiel2020}: $p_{\rm det} \leq 1 / n_{\rm init}$. Note that a classical random walk always has $p_{\rm det} = 1$; it reaches the target node in the end, but its hitting time may still be longer than that of a quantum walk since $p_{\rm th} \neq p_{\det}$.

\begin{figure}
\centering
\includegraphics[width=\linewidth]{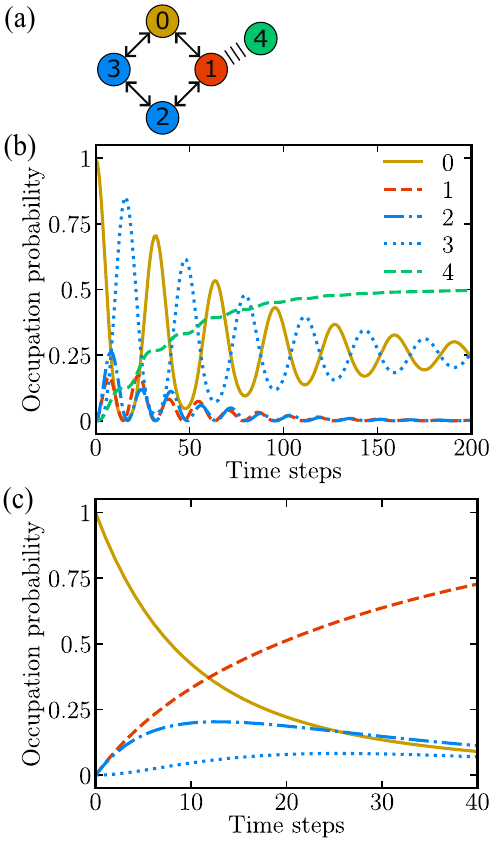}
\caption{Quantum and classical random walks on a four-node cyclic graph.
(a) The structure of the graph, with initial, target, and sink nodes marked following the same convention as in \figref{fig:adjacency_graph_matrix}.
(b) The dynamics of the QTRW on this graph.
(c) The dynamics of the CTQW on this graph.
\label{fig:dark}}
\end{figure}

In \figref{fig:dark}, we show the behaviour of classical random and quantum walks on a cyclic graph with four nodes. As can be seen in \figpanel{fig:dark}{a}, node 3 would be an equivalent initial state, and we thus have $p_{\rm det} = \frac{1}{2}$. The dark state formed by nodes 0 and 3 is the reason why \figpanel{fig:dark}{c} shows that the node occupation probability of the quantum walk displays oscillations for the two equivalent initial states (0 and 3) and ends with half of the energy in the target node and half evenly distributed between nodes 0 and 3.

%---------------------------------------------------------------------------------%
%---------------------------------------------------------------------------------%

%%%%%%%%%%%%%%%%%%%%%%%%%%%%%%%%%%%%%%%%%%%%%%%

\section{Methods}
\label{sec:Methods}

We apply machine learning to classify whether a CTRW or a CTQW yields an advantage on a given graph. In this section, we describe how we train neural networks to perform this classification and what neural-network architectures we use.

%%%%%%%%%%%%%%%%%%%%%%%%%%%%%%%%%

\subsection{Data generation}
\label{sec:DataGeneration}

To train the neural networks for their classification tasks, we first need to show the models many different examples of graphs where the classification is known. These examples are generated by numerical simulations, comparing the hitting times of CTRWs and CTQWs on a large number of graphs. Below, we explain how these simulations were carried out and how the graphs were generated. We also discuss how some data sets were pruned in order to improve training outcomes for the neural networks.

%%%%%%%%%%%%%%%%%%%%%%

\subsubsection{Generating graphs}
\label{sec:RandomGraphGeneration}

We consider three types of graphs: line graphs (the nodes form a line), cyclic graphs (the nodes form a closed loop), and a type of random graphs. The graphs were handled with NetworkX~\cite{team2014networkx}, a Python package for working with complex networks. For the first two types of graphs, we generated all possible configurations of initial and target placements for a given number of nodes.

For random graphs, we followed a recipe similar to the Erd\"osÐR\'enyi model for generating random graphs~\cite{erdos59a}, but we ensured that the resulting graph does not have nodes that are not connected to other nodes. In this graph-generation method, each edge has a fixed probability of being present or absent, independently of the other edges. We set the probability of creating edges between any two nodes to 0.05. This low probability makes the graphs less dense, which leads to a larger part of the data set containing graphs where a CTQW outperforms a CTRW.

%%%%%%%%%%%%%%%%%%%%%%

\subsubsection{Numerical simulations of random walks}
\label{sec:SimDynamicRWs}

We performed numerical simulations of the CTRW and CTQW for each generated graph. For the CTQW, we used the Python package QuTiP (Quantum Toolbox in Python)~\cite{Johansson2012, Johansson2013} to solve \eqref{eq:lindbladian}. From these simulations, we obtained the vector $p(t)$ of the occupation probabilities in all nodes as a function of time. The main information we extracted from this was the hitting time $\tau$ for the CTRW and the CTQW for each graph, which we used to classify the graph as yielding quantum advantage or not.

For the CTRWs, the simulations started with a classical particle in the initial node $i = 0$. For the CTQWs, we considered two different initial states: a quantum particle starting in the state $\ket{0}$ and a quantum particle starting in $\ket{T} = \cos \beta \ket{0} + e^{i\pi/4} \sin \beta \ket{n+1}$, with $\cos 2\beta = \frac{1}{\sqrt{3}}$ (see \secref{sec:Magic}).

%%%%%%%%%%%%%%%%%%%%%%

\subsubsection{Pruning data sets}
\label{sec:PruningDataSets}

In general, neural-network classifiers perform best when they can be trained on a data set containing roughly equal amounts of examples for each class they are to learn to classify~\cite{6469237}. In all the data sets we generated, we had more examples of graphs where the classical walk was faster than a quantum walk than vice versa. Before using these data sets for training, we therefore removed examples from the majority until we had an approximately equal number of examples showing classical and quantum advantages. Since the adjacency matrix of a given graph does not depend on vertex labeling, except for the initial and target node, a single graph can be used to generate multiple samples by shuffling the order for the nodes $i = 2, \ldots, n$.

In many cases, we noted that the difference between classical and quantum advantage was quite small, i.e., the two types of walks had almost the same hitting time on some graphs. Since neural networks in some cases have been shown to achieve better classification results when removing cases on the border between two classes~\cite{Goodfellow-et-al-2016}, we trained some neural networks on data sets where examples with similar hitting times had been removed. We also attempted the opposite pruning of data sets, removing instead the cases where the difference between quantum and classical walks was the largest (e.g., graphs where dark states and other factors made the hitting time of a quantum walk very long, or even infinite). Details of this these data-pruning methods, and the results from using them, are given in \appref{app:DataSimplifications}.

%%%%%%%%%%%%%%%%%%%%%%%%%%%%%%%%%

\subsection{Neural-network architectures}
\label{sec:ANN}

We tried out several neural-network architectures. Three main ones are presented here; results from other attempts, e.g., with a residual neural network (ResNet), were not promising.
First, we used a simple deep, fully connected (FC) feed-forward neural network~\cite{LeCun2015} with up to ten layers of neurons. We settled on using three layers after finding that deeper networks did not perform better. We also experimented with different widths of the network, i.e., the number of neurons in each layer, and settled on ten neurons, as wider networks than that did not perform better. The details of this architecture are shown in \tabref{table:dense_stucture}.

%---------------------------------------------------------------------------------%
\begin{table}
\begin{center}
\begin{tabular}{cc}
\hline
\hline
{Layer (type)} & {Output Shape} \\
\hline 
Input layer & $(n,n)$\\
Flattening \\
Dense & 10 \\
Dense & 10  \\ 
Dense & 10  \\ 
Dense & 2 \\
\hline
\end{tabular}
\end{center}
\caption{Definitions and shapes for the layers of the fully connected network.}
\label{table:dense_stucture}
\end{table}
%---------------------------------------------------------------------------------%

We also added convolutional layers before the fully connected ones since convolutional neural networks (CNNs)~\cite{Fukushima1980, LeCun2015, Dhillon2020} are particularly suited for classifying images~\cite{Krizhevsky2012} and have been used to classify graphs~\cite{Niepert2016, Tixier2017, Kawahara2017}. Our approach here differs from that in Refs.~\cite{Melnikov2019, Melnikov2020}, who used non-learnable convolutional filters to extract information about the connectivity of each node and the positions of the initial and target nodes, in that we make our convolutional filters fully learnable. We hoped that this added flexibility would have a positive impact on the performance of the classifier. The details of our CNN architecture are presented in \tabref{table:CNN_stucture}. %kernel size 3x3

The third network structure we study is that of Refs.~\cite{Melnikov2019, Melnikov2020}. There, the adjacency matrix is first fed into a sequence of convolutional layers with six filters (feature detectors), which are designed to detect how the initial and target nodes are connected to other nodes and how well-connected the other nodes in the graph are. Then there follows a filter that removes symmetric parts of the matrices, and $3 \times 3$ filters for finding relations between edges of the graph. The output is then compressed by further filters, flattened, and fed through two layers of fully connected neurons to yield the classification. For the full details of this architecture, we refer to the Methods section of Ref.~\cite{Melnikov2019}. Following the terminology introduced in Ref.~\cite{Melnikov2019}, we denote this architecture as a classical-quantum convolutional neural network (CQCNN).

%---------------------------------------------------------------------------------%
\begin{table}
\begin{center}
\begin{tabular}{cc}
\hline
\hline
{Layer (type)} & {Output Shape} \\
\hline 
Input layer & $(n,n)$ \\
Convolutional & $n$ \\
Convolutional & 10 \\
Convolutional & 10 \\
Convolutional & 10 \\
Dropout \\
Flattening \\
Dense & 10 \\
Dense & 10 \\ 
Dense & 10 \\ 
Dense & 2 \\
\hline
\end{tabular}
\end{center}
\caption{Definitions and shapes for the layers of the CNN. No padding is used in the kernel for the convolutional layers. The kernel size is (3,3). The dropout rate is 0.2.}
\label{table:CNN_stucture}
\end{table}
%---------------------------------------------------------------------------------%

In all architectures we investigate, we use the ReLu (rectified linear unit) activation function in all layers except the last one, where we instead use a softmax activation function to map our output to a probability distribution.

\subsection{Training and performance evaluation}
\label{sec:TrainingPerformance}

The neural networks were trained using the hyperparameters shown in \tabref{table:Hyperparameters}. The batch size (10) was fixed after testing a wide range of possibilities. The number of epochs (50) was chosen based on observed convergence. For comparison, Ref.~\cite{Melnikov2019} used a batch size of 3 and trained for 2000--3000 epochs. Learning rates spanning several orders of magnitude were tested before settling on the one used (0.001) in the results shown in \secref{sec:Results}.

To evaluate the performance of the trained models, we used four different performance metrics (accuracy $A$, precision $P$, recall $R$, and F1 score $F$)~\cite{bishop_2016} and the cross-entropy loss function on a test data set separate from the training data set. These performance metrics are defined as
\begin{align}
A &= \frac{TP + TN}{TP + TN + FP + FN}, \\
P &= \frac{TP}{TP + FP}, \\
R &= \frac{TP}{TP + FN}, \\
F &= \frac{2}{R^{-1} + P^{-1}} = \frac{2 TP}{2TP + FP + FN}, 
\end{align}
where $TP$ denotes true positive, $FP$ denotes false positive, $TN$ denotes true negative, and $FN$ denotes false negative. In this article, we always have only two classes, so these measures can focus on one of them in turn, denoting that as the positive and the other as the negative. We always averaged over ten training runs to obtain the numbers reported for these measures in \secref{sec:Results}. 

We evaluated the generalization ability of the trained networks by also testing them on data sets with a different number of nodes than in their training set. For these instances, the size of the network was adapted to the largest graph size that we wanted to test on, while the training was done on smaller graphs, where unconnected nodes were added to keep the total number of nodes in the graphs constant across training and test sets.

%---------------------------------------------------------------------------------%
\begin{table}
\begin{center}
\begin{tabular}{cc}
\hline
\hline
Hyperparameter & Value \\
\hline 
Learning rate &  0.001 \\
Batch size  & 10 \\ 
Number of epochs & 50 \\
\hline
\end{tabular}
\end{center}
\caption{Hyperparameters used when training the three types of neural networks.}
\label{table:Hyperparameters}
\end{table}
%---------------------------------------------------------------------------------%

%%%%%%%%%%%%%%%%%%%%%%%%%%%%%%%%%%%%%%%%%%%%%%%

\section{Results}
\label{sec:Results}

The behaviour of both CTRWs and CTQWs can vary significantly depending on the layout of the graph. In this section, we show results for fully connected and convolutional neural networks applied to the task of classifying whether there is an advantage for a quantum walk over a classical walk in cyclic, linear, and random graphs (see \secref{sec:RandomGraphGeneration}).

%%%%%%%%%%%%%%%%%%%%%%%%%%%%%%%%%

\subsection{Cyclic graphs}
\label{sec:ResultsCyclicGraphs}

In cyclic graphs, the behavior of a random walk is determined by the positions of the initial and target nodes (and the number of nodes). As outlined in \secref{sec:dark}, quantum walks in these graphs often exhibit quantum interference, which can yield dark states. From the symmetry argument in \secref{sec:dark}, it is clear that $p_{\rm det}$ is $\frac{1}{2}$ for all possible positions of the initial and target states if the graph has an odd number of nodes. If the graph has an even number of nodes, $p_{\rm det}$ will be 1 when the initial node is directly opposite the target node, and $\frac{1}{2}$ for all other cases.

%---------------------------------------------------------------------------------%
\begin{figure*}
\centering
\includegraphics[width = \linewidth]{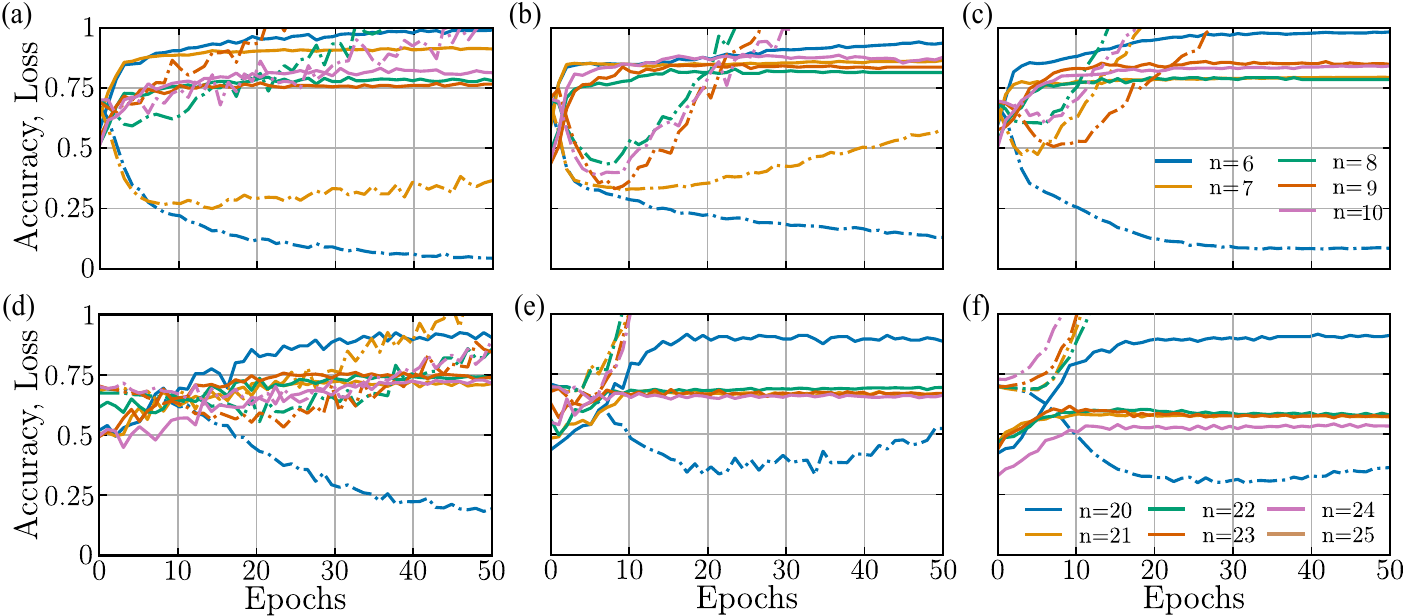} 
\caption{Results on cyclic graphs for the three neural-network architectures. The data set consists of all possible cyclic graphs for training and all possible cyclic graphs for testing. For 6 nodes, there are 15 different cyclic graphs; for 25 nodes, there are 300. Results for (a--c) graphs with 6--10 nodes from training on 6 nodes and (d-f) graphs with 20--25 nodes from training on 20 nodes. Solid lines show test accuracy and dash-dotted lines show test loss. The network architectures are (a, d) CQCNN, (b, e) CNN, and (c, f) FC.}
\label{fig:ResultsCyclic}
\end{figure*}
%---------------------------------------------------------------------------------%

In \figref{fig:ResultsCyclic}, we plot, for cyclic graphs with 6--10 and 20--25 nodes, the classification accuracy (share of data that was correctly classified) and loss of the fully connected and convolutional neural networks on a test set, as a function of number of epochs of training. For the results with 6--10 nodes, the networks were trained exclusively on training data with 6 nodes; similarly, the results for 20--25 nodes are for networks trained on training data with 20 nodes.

For the 6-node graphs that the networks were trained on, the classification accuracy reaches above \unit[98]{\%} for the CQCNN and the FC architectures; the accuracy is somewhat lower for the CNN architecture, but still high. As the size of the graph increases, the ability of the network to generalize decreases quickly, as shown most clearly by how the test loss increases in every case. Only for the small step to 7 nodes do the CQCNN and the CNN architectures manage to generalize decently.

For the 20-node graphs that the networks were trained on, the classification accuracy reaches about \unit[90]{\%} for all three architectures. The generalization ability, as shown by both accuracy and test loss, when scaling up to more than 20 nodes, is quite low for all three architectures. The CQCNN generalizes better than the other two architectures, but still displays a significant drop in accuracy when going from 20 to 21 nodes.

%%%%%%%%%%%%%%%%%%%%%%%%%%%%%%%%%

\subsection{Linear graphs}
\label{sec:ResultsLinearGraphs}

In linear graphs, we can make similar symmetry arguments as in Secs.~\ref{sec:dark} and \ref{sec:ResultsCyclicGraphs} to see that $p_{\rm det}$ for a CTQW is 1 for all configurations of initial and target nodes, except for the case with the target node being in the middle of a graph with an odd number of nodes, where the symmetry yields $p_{\rm det} = \frac{1}{2}$ for all placements of the initial node.

%---------------------------------------------------------------------------------%
\begin{figure*}
\centering
\includegraphics[width = \linewidth]{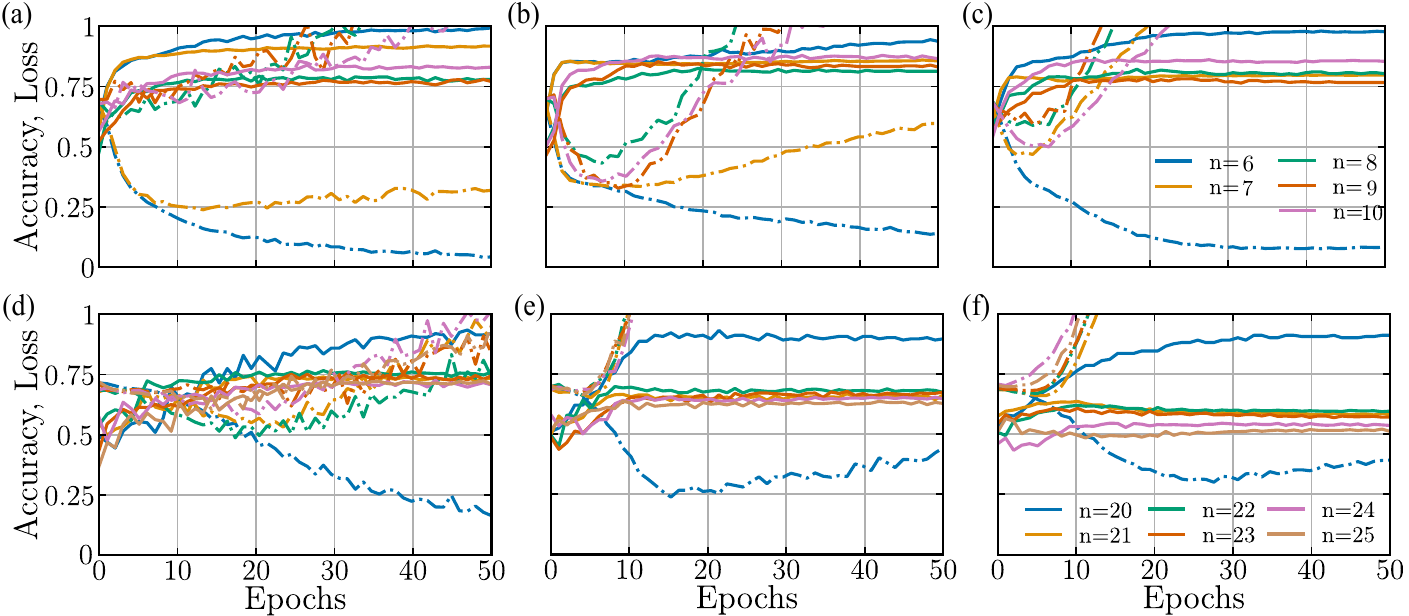} 
\caption{Results on linear graphs for the three neural-network architectures. The data set consists of all possible linear graphs for training and all possible linear graphs for testing. For 6 nodes, there are 15 different linear graphs; for 25 nodes, there are 300. Results for (a--c) graphs with 6--10 nodes from training on 6 nodes and (d-f) graphs with 20--25 nodes from training on 20 nodes. Solid lines show test accuracy and dash-dotted lines show test loss. The network architectures are (a, d) CQCNN, (b, e) CNN, and (c, f) FC.}
\label{fig:ResultsLinear}
\end{figure*}
%---------------------------------------------------------------------------------%

Just as we did for cyclic graphs in \figref{fig:ResultsCyclic}, we plot in \figref{fig:ResultsLinear} the classification accuracy and test loss of the fully connected and convolutional networks for two cases: test sets with 6--10 nodes for networks trained on 6 nodes, and test sets with 20--25 nodes for networks trained on 20 nodes.

The results in \figpanels{fig:ResultsLinear}{a}{c} for the networks trained on linear graphs with 6 nodes are almost identical to those for cyclic graphs in \figpanels{fig:ResultsCyclic}{a}{c}. The CQCNN and FC networks reach very high accuracy on the 6-node graphs, but all networks have large difficulties generalizing, particularly the FC architecture.

Also the results in \figpanels{fig:ResultsLinear}{d}{f}, for networks trained on linear graphs with 20 nodes, are very similar to those for cyclic graphs in \figpanels{fig:ResultsCyclic}{d}{f}. All three network architectures reach a classification accuracy of about \unit[90]{\%}. None of the three architectures display good generalization abilities. The CQCNN architecture generalizes better than the CNN architecture, which in turn does better than the FC architecture, but even the CQCNN accuracy drops from $\sim \unit[90]{\%}$ to $\sim \unit[75]{\%}$ when going from 20 to 21 or more nodes.

%%%%%%%%%%%%%%%%%%%%%%%%%%%%%%%%%

\subsection{Random graphs}
\label{sec:ResultsRandomGraphs}

We performed our most extensive testing of different classification methods on random graphs since they encompass a much greater variety of possible patterns than cyclic and linear graphs for neural networks to pick up on and exploit.

%%%%%%%%%%%%%%%%%%%%%%

\subsubsection{Principal component analysis}
\label{sec:PCA}

\begin{figure}
\centering
\includegraphics[width = \linewidth]{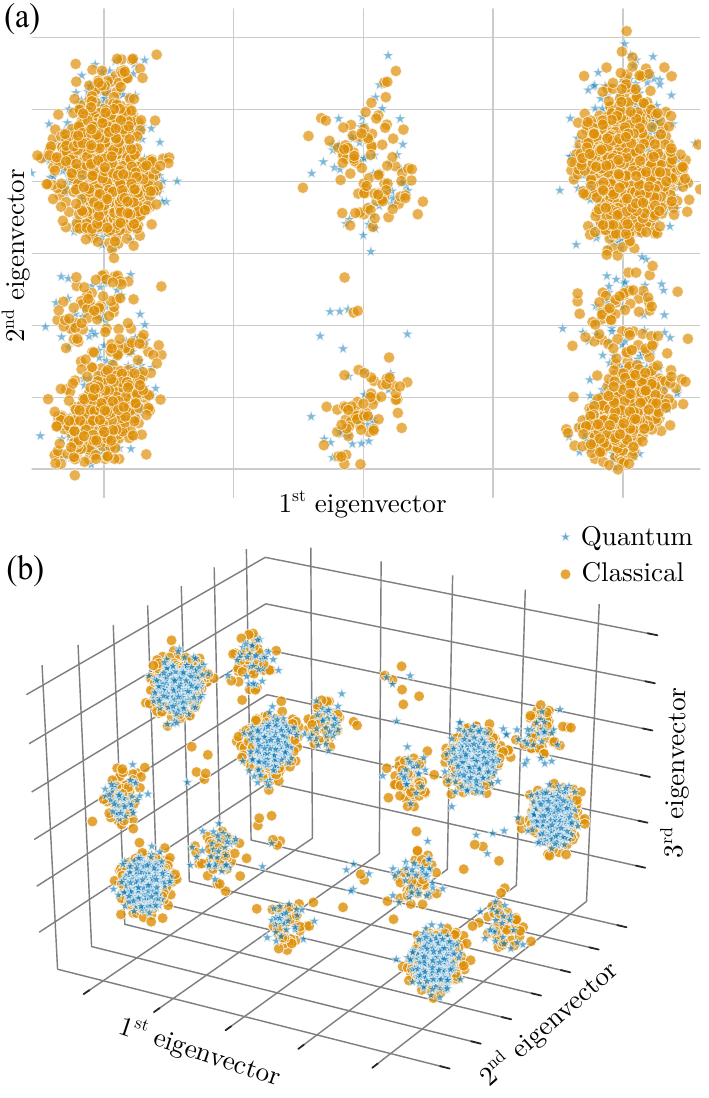}
\caption{Data from 5000 random graphs with 20 nodes sorted along the first (a) two and (b) three directions from PCA. The cases where the CTRW is faster are marked as yellow dots. The cases where the CTQW is faster are shown as blue stars.}
\label{fig:pca_20Random}
\end{figure}

Before applying the full force of neural networks to a classification problem, it is usually good to check if a simpler method performs well. Here, we use principal component analysis (PCA)~\cite{Abdi2010, Jolliffe2016} to see if the high-dimensional problem we are dealing with can be reduced to a lower-dimensional one where classification becomes easier. In \figref{fig:pca_20Random}, we show the result of PCA performed on the training data set for random graphs with 20 nodes. We see that when reducing the problem to two or three dimensions, there is no clear boundary between the cases where quantum or classical random walks are faster. This result is indicative of the complexity of our classification task and motivates the application of neural networks to this task. The results for three-dimensional PCA do indicate that there is some sort of structure in the data, which a neural network could pick up on and use.

%---------------------------------------------------------------------------------%
%---------------------------------------------------------------------------------%

%%%%%%%%%%%%%%%%%%%%%%

\subsubsection{Performance of neural networks on random graphs}
\label{sec:PerformanceNNRandomGraphs}

%---------------------------------------------------------------------------------%
\begin{figure*}
\centering
\includegraphics[width = \linewidth]{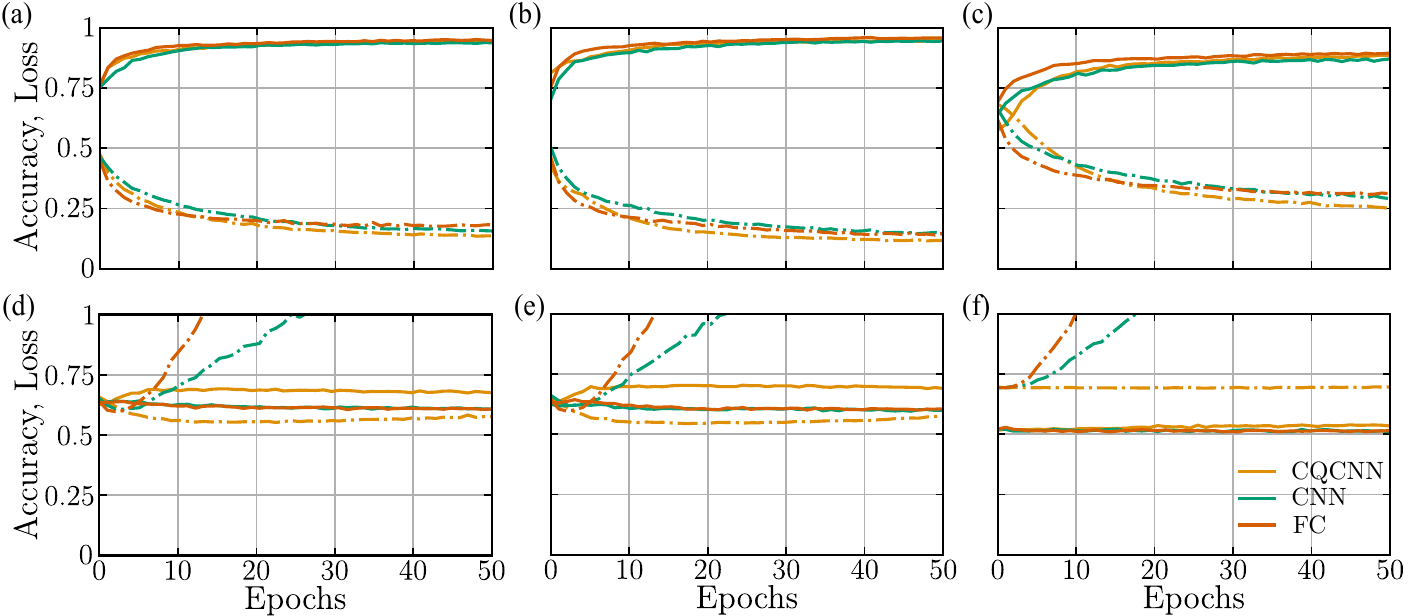}
\caption{Results for the three neural-network architectures trained and tested on random graphs with (a--c) 6 and (d--f) 20 nodes. In panels (a) and (d), the networks compare classical and quantum walks. In panels (b) and (e), the networks compare classical random walks to quantum walks that start in $\ket{T}$. In panels (c) and (f), the networks compare quantum walks that start in the initial node to quantum walks that start in $\ket{T}$. The data sets used are derived from a set of 5000 random graphs. After removing some graphs to have an equal amount in each class, 2620--4716 graphs remain, depending on which number of nodes and which two classes are considered; they are split 80/20 into a training set and a test set. Solid lines show test accuracy and dashed-dotted lines show test loss.
}
\label{fig:ResultsRandom}
\end{figure*}
%---------------------------------------------------------------------------------%

The results of testing our three neural-network architectures on 6- and 20-node random graphs are shown in Figs.~\figpanelNoPrefix{fig:ResultsRandom}{a} and \figpanelNoPrefix{fig:ResultsRandom}{d}. For 6 nodes, all three architectures perform equally well, reaching classification accuracies above \unit[90]{\%}. For 20 nodes, all networks fail to reach even \unit[70]{\%} accuracy. The CQCNN does best here; unlike the CNN and the FC, it does not overfit (displaying an increasing test loss).

%---------------------------------------------------------------------------------%
\begin{table}
\begin{center}
\begin{tabular}{cccc}
\hline
\hline
No.~of nodes & Precision & Recall & F1 score \\
\hline 
    6 &  [0.96, 0.96] & [0.99, 0.87] & [0.98, 0.92] \\
    20  &  [0.73, 0.53] & [0.71, 0.55] & [0.72, 0.54] \\
\hline
\end{tabular}
\end{center}
\caption{Precision, recall, and F1 score for the CQCNN classifying between classical (first number) and quantum (second number) random walks on random graphs with 6 and 20 nodes.}
\label{tab:cq_recall}
\end{table}
%---------------------------------------------------------------------------------%

To further quantify the performance of the CQCNN, we provide, in Table~\ref{tab:cq_recall}, the additional neural-network performance measures of precision, recall, and F1 score (see definitions in \secref{sec:TrainingPerformance}). All these measures are calculated at the end of training, i.e., after 50 epochs in \figref{fig:ResultsRandom}. For the 6-node case, all scores are high. For the 20-node case, we can compare with results in Ref.~\cite{Melnikov2019}: we find lower precision and recall for the graphs with classical advantage, but higher precision and recall for the graphs with quantum advantage. Since the network architecture is the same here, we believe these differences may be due to how we have selected the training and test data to contain equal amounts of examples with classical and quantum advantages.

We refer to \appref{app:DataSimplifications} for additional results obtained by pruning the data sets from either cases that were hard to distinguish or from cases that were easy to distinguish. Neither method yielded any improvement over the results shown in this section.

%%%%%%%%%%%%%%%%%%%%%%

\subsubsection{Classifying quantum walks starting from a superposition state}
\label{sec:ResultsSuperposition}

Finally, we also consider classification where the quantum walk starts in the $\ket{T}$ superposition state defined in \eqref{eq:T_state_ket}. We compare such quantum walks to both classical walks and quantum walks starting fully in the initial node.

In Figs.~\figpanelNoPrefix{fig:ResultsRandom}{b} and \figpanelNoPrefix{fig:ResultsRandom}{e}, we show classification results for all three neural-network architectures on 6- and 20-node random graphs, where the two classes are classical random walks and quantum walks starting in $\ket{T}$. For 6 nodes, all three architectures basically perform equally well, reaching classification accuracies above \unit[90]{\%}, very similar to the results in \figpanel{fig:ResultsRandom}{a} for classification between classical walks and quantum walks starting fully in the initial state. For 20 nodes, the results are similar to those in \figpanel{fig:ResultsRandom}{d}; the CQCNN reaches about \unit[70]{\%} accuracy while the CNN and FC architectures stop around \unit[60]{\%} accuracy and display an increasing test loss.

%---------------------------------------------------------------------------------%
\begin{table}
\begin{center}
\begin{tabular}{cccc}
\hline
\hline
No.~of nodes & Precision & Recall & F1 score \\
\hline 
    6 & [0.94, 0.95] & [0.97, 0.89] & [0.96, 0.92] \\
    20 & [0.73, 0.53] & [0.75, 0.50] & [0.74, 0.52] \\
\hline
\end{tabular}
\end{center}
\caption{Precision, recall, and F1 score for the CQCNN classifying between classical random walks (first number) and quantum walks starting in $\ket{T}$ (second number) on random graphs with 6 and 20 nodes. We also obtained this data for 25-node random graphs; the numbers were almost exactly the same as for 20 nodes.}
\label{tab:cT_recall}
\end{table}
%---------------------------------------------------------------------------------%

In Table~\ref{tab:cT_recall}, we provide the additional neural-network performance measures of precision, recall, and F1 score for the CQCNN results (at the end of training) in Figs.~\figpanelNoPrefix{fig:ResultsRandom}{b} and \figpanelNoPrefix{fig:ResultsRandom}{e}. For both the 6-node and 20-node cases, the results are almost the same as those in \tabref{tab:cq_recall} for classification between classical random walks and quantum walks starting fully in the initial state.

In Figs.~\figpanelNoPrefix{fig:ResultsRandom}{c} and \figpanelNoPrefix{fig:ResultsRandom}{f}, we compare the ability of the three neural-network architectures to distinguish between quantum walks starting in $\ket{T}$ and quantum walks starting in the initial node, on 6- and 20-node random graphs. This appears to be a harder task than differentiating classical walks and different quantum walks. For the small 6-node graphs, all three architectures do fairly well, reaching classification accuracies close to \unit[90]{\%}. However, for the 20-node graphs, none of the three networks is able to reach an accuracy significantly removed from \unit[50]{\%}, which is the same accuracy as random guessing.

%---------------------------------------------------------------------------------%
\begin{table}
\begin{center}
\begin{tabular}{cccc}
\hline
\hline
No.~of nodes & Precision & Recall & F1 score \\
\hline 
    6 & [0.83, 0.93]& [0.92, 0.85] & [0.87, 0.89] \\
    20  & [0.49, 0.47] & [0.60, 0.35] & [0.54, 0.40] \\
\hline
\end{tabular}
\end{center}
\caption{Precision, recall, and F1 score for the CQCNN classifying between quantum walks starting in the initial node (first number) and quantum walks starting in $\ket{T}$ (second number) on random graphs with 6 and 20 nodes.}
\label{tab:qT_recall}
\end{table}
%---------------------------------------------------------------------------------%

In Table~\ref{tab:qT_recall}, we provide the additional neural-network performance measures of precision, recall, and F1 score for the CQCNN results (at the end of training) in Figs.~\figpanelNoPrefix{fig:ResultsRandom}{b} and \figpanelNoPrefix{fig:ResultsRandom}{e}. Just as the accuracy decreased compared to classification involving classical and quantum examples, these additional measures show lower values than those in Tables~\ref{tab:cq_recall} and \ref{tab:cT_recall}.

%%%%%%%%%%%%%%%%%%%%%%%%%%%%%%%%%%%%%%%%%%%%%%%

\section{Conclusion and outlook}
\label{sec:ConclusionOutlook}

We have studied the use of neural networks to detect a quantum walk has an advantage over a classical random walk. We focused on three different neural-network architectures: a fully connected neural network, a convolutional neural network (followed by fully connected layers), and the architecture of Ref.~\cite{Melnikov2019} --- a CNN with pre-defined filters. We tested the performance of these networks on classification of linear, cyclic, and random graphs with up to 25 nodes. In some cases, we tested classification of quantum walks starting in the $\ket{T}$ superposition state.

Our results indicate that the fully connected and convolutional networks did not improve on the performance of the CQCNN proposed in Ref.~\cite{Melnikov2019}. For many cases, the difference in performance between the three networks tested is small, but the CQCNN performs better on some of the most difficult test sets --- 20-node random graphs. We note that our careful procedure for selecting large and evenly distributed sets of training data seems to have improved performance on classifying the cases with a quantum advantage compared to Ref.~\cite{Melnikov2019}, without changing the network architecture itself.

However, it is also clear that none of the three architectures performs particularly well for these most challenging tests, neither in terms of accuracy, precision, recall, nor F1 score, although the results are clearly better than chance. The classification problem itself appears to be a tough nut to crack. This level of difficulty is also indicated by the principal component analysis we performed for 20-node random graphs.

We further see from our tests that the ability of all three networks to generalize to larger graphs after training on smaller graphs is quite limited, even if the graphs are highly regular (cyclic or linear). Distinguishing between quantum walks starting in different types of initial states is also clearly more challenging than distinguishing between classical random walks and quantum walks.
 
Overall, our results suggest that neural networks can be used to tackle the challenging problem of detecting advantages of quantum walks over classical random walks, and provide insights into the behavior of quantum walks on complex graph structures. However, the performance of the neural networks needs to be further improved for them to become truly useful.

One possible reason for the overall lackluster performance of the studied networks is that the input data (the adjacency matrices of the graphs) is sparse. In general, sparse input data can be challenging for neural networks to train on~\cite{Hastie2015, Petrini2022}. It may be that the somewhat better performance of the CQCNN (compared to the FC and convolutional neural networks) is due, at least in part, to the first layers there making the data more dense.

We see several ways to build further on the work and results reported here. In the search for better performance, one could attempt much larger versions of the architectures we considered here or other hyperparameters beyond what we explored in this article. Another possibility is to test other neural-network architectures, e.g., various versions of graph neural networks~\cite{Zhou2020}, which have emerged in recent years as a way to tackle machine-learning problems that can be formulated in terms of graphs, or other networks better suited to sparse input data. In such a search for other architectures, it is desirable to find one that is good at generalization, i.e., that can be trained on graphs of a certain size and then manages to also classify graphs of other sizes.

After further performance improvements, or perhaps already with the level of performance we have reached in this work, it will be interesting to extract what features the neural network learns to classify. For this task, methods such as gradient-weighted class activation mapping (Grad-CAM)~\cite{Selvaraju2020} can be used. As we see it, the ultimate goal of this line of research is not to find a neural network that can identify quantum advantage, but to be able to learn from such a network what features in a system determine whether there is a quantum advantage or not. Such insights, or architectures for neural networks, may find applications not only for random walks, but for other information-processing methods where a quantum advantage is sought or in fields like quantum transport.

%%%%%%%%%%%%%%%%%%%%%%%%%%%%%%%%%%%%%%%%%%%%%%%

\begin{acknowledgments}

We thank Alexey Melnikov for sharing the code from Ref.~\cite{Melnikov2019}. We acknowledge useful discussions with Evert van Nieuwenburg, David Fitzek, Shahnawaz Ahmed, and Baptiste Cavarec. 

We acknowledge support from the Knut and Alice Wallenberg Foundation through the Wallenberg Centre for Quantum Technology (WACQT).
AFK is also supported by the Swedish Research Council (grant number 2019-03696) and the Swedish Foundation for Strategic Research (grant numbers FFL21-0279 and FUS21-0063).

\end{acknowledgments}

\section*{Code availability}

The code for producing the results is provided upon reasonable request.
All calculations are performed with Numpy~\cite{harris_array_2020}, Tensorflow~\cite{tensorflow2015-whitepaper}, and Keras~\cite{chollet2015keras}, and all the plots are generated with the help of Matplotlib~\cite{hunter_matplotlib_2007}.

%%%%%%%%%%%%%%%%%%%%%%%%%%%%%%%%%%%%%%%%%%%%%%%

\appendix

%%%%%%%%%%%%%%%%%%%%%%%%%%%%%%%%%%%%%%%%%%%%%%%

\section{Pruning data sets based on distinguishability}
\label{app:DataSimplifications}

%---------------------------------------------------------------------------------%
\begin{figure}
\centering
\includegraphics[width = 0.9\linewidth]{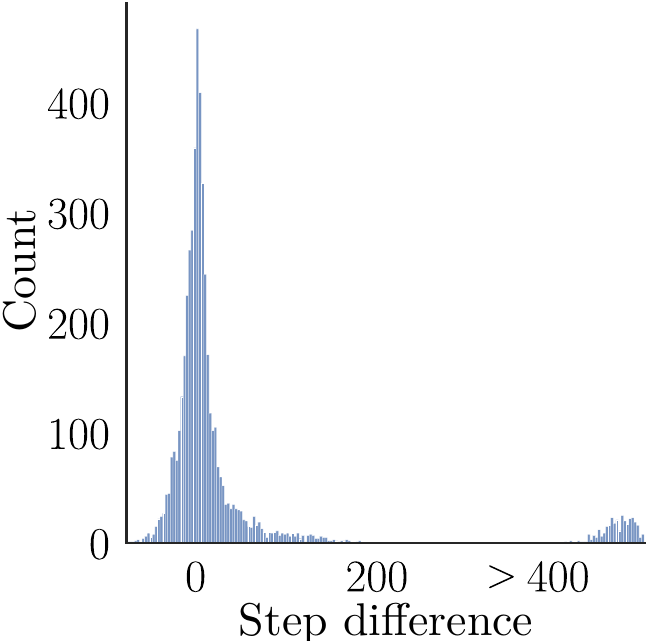}
\caption{The difference in the number of time steps $dt = 0.01$ required to reach the threshold for quantum and classical walkers on 20-node random graphs. The data set consists of 5000 graphs.}
\label{fig:data_random20}
\end{figure}
%---------------------------------------------------------------------------------%

As mentioned in \secref{sec:PruningDataSets}, we also investigated the performance of our neural networks when the hard-to-distinguish or easy-to-distinguish cases were removed from the training data. The motivation for doing so can be seen from the plot in \figref{fig:data_random20}, which shows the distribution of the difference in number of time steps required to reach the threshold for classical and quantum walkers on the data set of 5000 random 20-node graphs that was used in \secref{sec:ResultsRandomGraphs}. The plot shows that there is a cluster of hard-to-distinguish cases with time-step difference close to zero, and a smaller cluster of easy-to-distinguish cases with time-step difference well above 400. In some of the latter cases, dark states (see \secref{sec:dark}) lead to the quantum walk never reaching the threshold at all.

%---------------------------------------------------------------------------------%
\begin{figure}
\centering
\includegraphics[width = \linewidth]{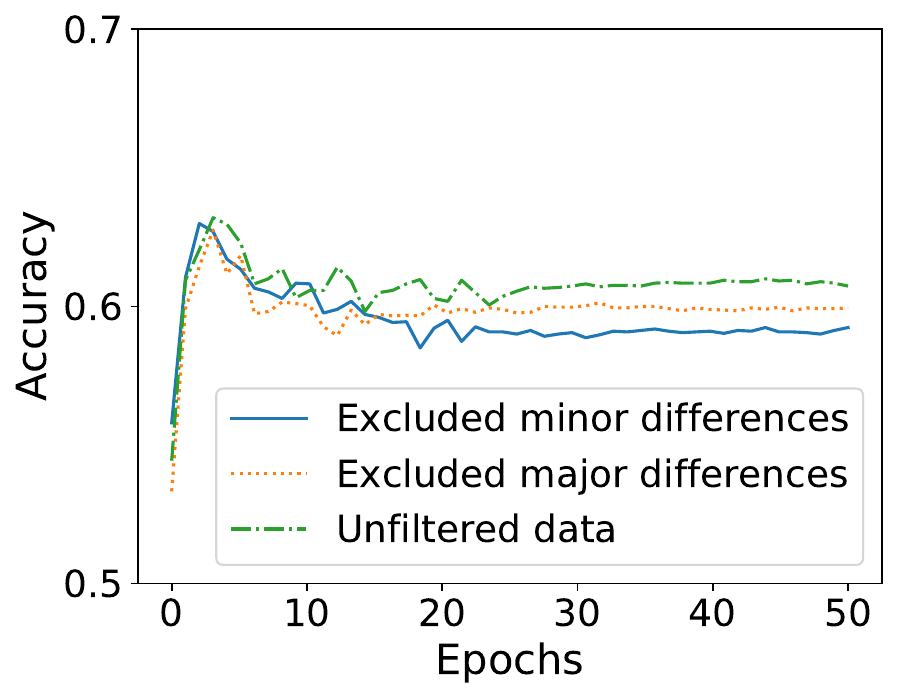}
\caption{Evaluation of the effects of omitting minor variations and significant step differences. Note that the scale of the y axis is different from other similar figures in the main text.}
\label{fig:random20DNN_Dis}
\end{figure}
%---------------------------------------------------------------------------------%

After removing some graphs to have an equal number of cases with classical and quantum advantage, we split this data set 80/20 into a training set and a test set. We then trained an FC neural-network architecture on three versions of this training set: one using all the data, one removing cases with a time-step difference of 5 or less, and one removing cases with a time-step difference of 200 or more. These three trained networks were then all tested on the same, full test set. 

The resulting accuracy as a function of the number of training epochs is shown in \figref{fig:random20DNN_Dis}. It turned out that removing some cases from the training data did not improve performance for the network. Removing the cases with a major difference slightly decreased performance; removing the cases with minor differences resulted in a somewhat clearer decrease in accuracy.

From these results, we draw two conclusions. Firstly, the cases with major differences are not that essential for training (the network mostly learns to understand them anyway), but their presence in the training set does not appear to be a complication for the network. Secondly, cases with small differences do not confuse the network during training, but rather need to be present in the data set for the network to learn something about how to distinguish them.

%%%%%%%%%%%%%%%%%%%%%%%%%%%%%%%%%%%%%%%%%%%%%%%

\bibliography{ReferencesQuantumWalks}

\end{document}